\documentstyle[12pt]{article}
\begin{document}

\newcommand{\be}{\begin{equation}}
\newcommand{\ee}{\end{equation}}
\newcommand{\ba}{\begin{eqnarray}}
\newcommand{\ea}{\end{eqnarray}}
\newcommand{\ct}{\cite}

\begin{center}
{\bf NEW PHASE STRUCTURE OF THE NAMBU -- JONA - LASINIO MODEL 
AT NONZERO CHEMICAL POTENTIAL}

\vspace*{0.3cm}
K.G. Klimenko$^{\dagger}$, A.S. Vshivtsev$^{\ddagger}$,

\vspace{0.2cm}
Institute for High Energy Physics, 142284, Protvino, Moscow region, Russia

\vspace{0.1cm}

E-mail:~$^\dagger$ kklim@mx.ihep.su, ~$^\ddagger$ alexandr@vvas.msk.ru 

\end{center}

\vspace{0.4cm}
It is shown that in the Nambu -- Jona - Lasinio model at nonzero
chemical potential there are two different
massive phases with spontaneously broken chiral symmetry. In one of them
particle density is identically zero, in another phase it is not equal to zero.
The transition between phases is a phase transition of the second order.

\vspace*{0.6cm}
This is an english translation of the article, published in Russian journal
Pis'ma Zh. Eksp. Teor. Fiz. (JETP Letters) \ct{100}.

It is well known that Nambu -- Jona - Lasinio (NJL) model \ct{1,2} is perfect
laboratory for the investigations of low energy region in QCD. The subject of 
special interest is the spontaneous breaking of chiral symmetry. In the 
framework of NJL model this phenomenon was studied at nonzero temperature and
density \ct{3,4}, in the presence of external fields \ct{5}, at nonzero
curvature and nontrivial topology of space - time \ct{6}.

In the present article the phase structure of NJL model at nonzero chemical
potential $\mu$ is considered. In contrast with earlier papers \ct{3,4} we
have found new massive phase of the model. The NJL Lagrangian has the form:
\be
L=\sum^{N}_{k=1}\bar\psi_k i\hat\partial\psi_k
+\frac{G}{2N}
[(\sum^{N}_{k=1}\bar\psi_k\psi_k)^2
+(\sum^{N}_{k=1}\bar\psi_k i\gamma_5\psi_k)^2].
\ee
It contains $N$ four - components Dirac fields $\psi$. So one can use $1/N$ - expansion
technique. Eq. (1) is invariant under continuous chiral transformations:
\be
\psi_k\to e^{i\theta\gamma_5}\psi_k (k=1,...,N).
\ee

First of all let us recall some well - known vacuum properties  of the theory (1)
at $\mu=0$. The introduction of an auxiliary Lagrangian
\be
\tilde L=\bar\psi i\hat\partial\psi-
\bar\psi (\sigma_1+i\sigma_2\gamma_5)
\psi-\frac{N}{2G}(\sigma^2_1+\sigma^2_2)
\ee
greatly facilitates the problem under consideration. (In (3) and other formulae
below we have omitted the fermionic index $k$ for simplicity.) On the equations of 
motion for auxiliary bosonic fields $\sigma_{1,2}$ the theory (3) is equivalent
to the (1) one.

From (3) it follows in the leading order of $1/N$ - expansion:
$$
\exp (i S_{eff}(\sigma_{1,2})) =
\int D\bar\psi D\psi \exp(i\int\tilde L d^4x),
$$
where
\be
\frac{1}{N}S_{eff}(\sigma_{1,2})=-
\int d^4x\frac{\sigma_1+\sigma^2_2}{2G}-i\ln\det
(i\hat\partial-\sigma_1-i\gamma_5\sigma_2).
\ee
Supposing that in this formula $\sigma_{1,2}$ are not dependent from 
the space - time points we have by definition:
\be
S_{eff}(\sigma_{1,2})=-
V_{eff}(\sigma_{1,2})\int d^4 x,
\ee
where $(\Sigma=\sqrt{\sigma^2_1+\sigma^2_2})$:
\be
\frac{1}{N}V_{eff}(\sigma_{1,2})=
\frac{\Sigma^2}{2G}+2i
\int \frac{d^4p}{(2\pi)^4}\ln
(\Sigma^2-p^2)\equiv \frac{1}{N}V_0(\Sigma).
\ee

Introducing in (6) Euclidean metrics $(p_0\to ip_0)$ and cutting off the range
of integration $(p^2\leq \Lambda^2)$, we obtain:
\ba
\frac{1}{N}V_0(\Sigma)&=&
\frac{\Sigma^2}{2G}-
\frac{1}{16\pi^2}\Biggl\{
\Lambda^4\ln
\left(1+\frac{\Sigma^2}{\Lambda^2}\right)+\Lambda^2\Sigma^2-\Biggr.\nonumber \\
&-&\Sigma^4\ln \left(1+\frac{\Lambda^2}{\Sigma^2}\right)\Biggl.\Biggr\}.
\ea
The stationary equation for the effective potential (7) has the form:
\be
\frac{\partial V_0(\Sigma)}{\partial\Sigma}=0=
\frac{N\Sigma}{4\pi^2}
\Biggl\{\frac{4\pi^2}{G}-\Lambda^2+\Sigma^2 \ln
\left (1+\frac{\Lambda^2}{\Sigma^2}\right)\Biggr\}\equiv
\frac{N\Sigma}{4\pi^2}F(\Sigma).
\ee
Now one can easily see that at $G<G_c=4\pi^2/\Lambda^2$ eq. (8) has no 
solutions apart from $\Sigma=0$. Hence, in this case fermions are massless,
and chiral invariance (2) is not broken.

If $G>G_c$, then Eq. (8) has one nontrivial solution $\Sigma_0(G,\Lambda) \not = 0$
such that $F(\Sigma_0)=0$. In this case $\Sigma_0$ is a point of global
minimum for the potential $V_0(\Sigma)$. This mean the spontaneous breaking of 
the symmetry (2) takes place. Moreover, fermions acquire mass $M\equiv\Sigma_0 (G,\Lambda)$.

In the following at $G>G_c$ we shall use the fermionic mass $M$ as an 
independent parameter of the theory. Three quantities $G$, $M$ and $\Lambda$ 
are connected by the Eq. (8). In terms of $M$ and $\Lambda$ the effective
potential $V_0(\Sigma)$ has an equivalent form:
\ba
\frac{16\pi^2}{N}V_0(\Sigma)=
\Sigma^2\Lambda^2&-&2\Sigma^2M^2\ln
(1+\Lambda^2/M^2)-\nonumber \\
-\Lambda^4\ln (1+\Sigma^2/\Lambda^2)&+&
\Sigma^4\ln (1+\Lambda^2/\Sigma^2)
\ea

Let us now imagine that $\mu>0$ and temperature $T \not =0$. In this case one can
obtain effective potential $V_{\mu T}(\Sigma)$ if the mesure of integration in
(6) is transformed as 
$$
\int\frac{dp_0}{2\pi}\to iT\sum^{\infty}_{n=-\infty},\,\,
p_0\to i\pi T(2n+1)+\mu.
$$
Summing there over $n$ \ct{7}, we have:
\ba
\frac{1}{N}V_{\mu T}(\Sigma)&=&\frac{1}{N}V_0(\Sigma)-2T
\int\frac{d^3p}{(2\pi)^3}\ln\left\{[1+ \right.\nonumber \\
+\exp (-\beta(\sqrt{\Sigma^2+\bar p^2}&+&\mu))]
[1+\exp (-\beta (\sqrt{\Sigma^2+\bar p^2}-\mu))]\left.\right\},
\ea
where $\beta=1/T$, and $V_0(\Sigma)$ is presented in (7-9). Since we are going 
to study phase structure of NJL model at $\mu>0$ and $T=0$ only, one should
direct the temperature to the zero in the above equation. From this we have
corresponding effective potential $V_\mu (\Sigma)$:
\ba
\frac{1}{N}V_{\mu}(\Sigma)=\frac{1}{N}V_0(\Sigma)&-&
\frac{\Theta(\mu-\Sigma)}{16\pi^2}
\Biggl\{\frac{10}{3}\mu(\mu^2-\Sigma^2)^{3/2}- \Biggr.\nonumber \\
-2\mu^3\sqrt{\mu^2-\Sigma^2}&+&\Sigma^4\ln
~[~(~\mu+\sqrt{\mu^2-\Sigma^2}~)^2/\Sigma^2~]\Biggl.\Biggr\}.
\ea
The function (11) is symmetric under transformation $\Sigma \to -\Sigma$, so
it is sfficient to learn the situation with $\Sigma\geq 0$ only.

In the present paper we shall consiger the case $G>G_c$ only. From (9) and (11)
one can find stationary equation for the effective potential
($M,\Lambda$ are free parameters):
\ba
&&\frac{\partial V_\mu(\Sigma)}{\partial\Sigma}=0=
\frac{N\Sigma}{4\pi^2}
\Biggl\{\Sigma^2\ln
\left [1+\frac{\Lambda^2}{\Sigma^2}\right]-M^2\ln
\left (1+\frac{\Lambda^2}{M^2}\right)\Biggr.\nonumber\\
&&+\Theta (\mu-\Sigma)~[~2\mu
\sqrt{\mu^2-\Sigma^2}-2\Sigma^2\ln
((\mu+\sqrt{\mu^2-\Sigma^2}~)/\Sigma)]\Biggl.\Biggr\}.
\ea

Let $\mu<M$ and $M<<\Lambda$. Then at $\Sigma>\mu$ eq. (12) coincides with 
stationary equation for the potential $V_0(\Sigma)$ (9). Hence, in this case
the single solution of (12) is $\Sigma_1=M$. At $\Sigma<\mu$ eq. (12) takes the form:
\ba
\Sigma f_\mu (\Sigma)\equiv \Sigma\Biggl\{2\mu
\sqrt{\mu^2-\Sigma^2}-M^2\ln 
\left(1+\frac{\Lambda^2}{M^2}\right)+\Biggr.\nonumber \\
+\Sigma^2\ln[(\Sigma^2+\Lambda^2)/(\mu+\sqrt{\mu^2-\Sigma^2})^2]\Biggl.\Biggr\}=0.
\ea
It is easily to show that at $M<<\Lambda$ the function $f_\mu (\Sigma)$ is
monotonically increasing one over variable $\Sigma\in [0,\mu]$. Obviously, that
$f_\mu (\mu)<0$ at $\mu< M$, 
$f_\mu (\mu)>0$ at $\mu >M$ and $f_\mu (\mu)=0$ at $\mu= M$. Consequently,
at $\mu<M$ the single solution of eq. (13) is the point $\Sigma_2=0$. So,
at $\mu<M$ the potential (11) has two stationary points:
$\Sigma_1=M$ and $\Sigma_2=0$. At $\Sigma_2=0$ there is a local maximum of the
potential. But the point $\Sigma_1=M$ is the point of global minimum for 
$V_\mu(\Sigma)$. Hence, at $\mu<M$ the chiral symmetry of the NJL model is
spontaneously broken and fermions have mass $M$.

Suppose now that $M<\mu<\mu_{1c}(M)$, where
\be
\mu_{1c}(M)
=\left[\frac{M^2}{2}\ln \left(1+\frac{\Lambda^2}{M^2}\right)\right]^{1/2}.
\ee
It is evident that for such values of the chemical potential $f_\mu(0)<0$ 
and $f_\mu(\mu)>0$. So the monotonically increasing function $f_\mu (\Sigma)$
(remember $M<<\Lambda$) crosses $\Sigma$ axis in the single point
$\Sigma_3(\mu, M)$, which has properties:
\ba
\Sigma_3 (\mu, M)&\to& M\;\;\; \mbox{при} \;\;\; \mu\to M\nonumber \\
\Sigma_3 (\mu, M)&\to& 0\;\;\; \mbox{при} \;\;\; \mu\to \mu_{1c}(M).
\ea
We have to remark, that in the case under consideration the point $\Sigma_1=M$
is no more the solution of the eq. (12), but $\Sigma_2=0$ is the local maximum
of the potential (11) as before. So at $M<\mu<\mu_{1c}(M)$ there is 
a spontaneous breaking of the symmetry in the NJL model, and fermions acquire mass $\Sigma_3(\mu, M)$.

At $\mu>\mu_{1c}(M)$ eqs. (12), (13) have the solution $\Sigma_2=0$ only.
So at $\mu>\mu_{1c}(M)$ we have massless phase A of the theory. Here the 
chiral symmetry is not broken. In the critical point $\mu_{1c}(M)$ there is
a phase transition of the second order from the massive phase to the massless one,
because the order parameter -- fermionic
mass -- is a continuous function, when $\mu = \mu_{1c}(M)$ (see (15)).

Now we are going to prove that indeed there are two different massive
phases of the theory. The first one -- the phase B -- is situated, when
$\mu < M$ and the second one -- the phase C -- is realized, when $M<\mu<\mu_{1c}(M)$.
So we shall try to investigate the thermodynamic potential (TP) $\Omega(\mu)$ of the NJL system at
the point $\mu=\mu_{2c}(M) \equiv M$. Let
\be
\Omega (\mu)\equiv
\left\{
\begin{array}{lcl}
\Omega_B(\mu),\;\;\;\mbox{при} \;\;\; \mu<M\nonumber \\
\Omega_C(\mu),\;\;\;\mbox{при} \;\;\; M<\mu<\mu_{1c}(M).\end{array}
\right.
\ee
Recall, that TP is the value of the effective potential at the point of global
minimum. Hence, $\Omega_B(\mu)=V_\mu(M),\;\; \Omega_C(\mu)=V_\mu
(\Sigma_3(\mu,M))$. Now let us calculate derivatives of $\Omega_B(\mu)$ and
$\Omega_C(\mu)$ at the point $\mu=M$. First of all one should note that from (15)
it follows $\Omega_B(M)=\Omega_C(M)$. From (9-11) we have
\ba
\Omega_B(\mu)=V_\mu(M)=V_0(M)=
\frac{N}{16\pi^2}
\left\{M^4-\right.\nonumber\\
[-0.2cm]\\
-M^4\ln (1+\Lambda^2/M^2)-\Lambda^4\ln (1+M^2/\Lambda^2)\left.\right\}.\nonumber
\ea
It is clear, that $\Omega_B(\mu)$ is $\mu$ independent function, that is why all
its derivatives over $\mu$ identically equel to zero in the region  $\mu< M$,
including the case $\mu\to M_-$. The first derivative of the $\Omega_c(\mu)$ is:
\be
\frac{d\Omega_c(\mu)}{d\mu}=
\Biggl\{
\frac{\partial V_\mu (\Sigma)}{\partial\mu}+
\frac{\partial V_\mu (\Sigma)}{\partial\Sigma}
\frac{\partial \Sigma}{\partial\mu}
\Biggr\}
\Biggl|_{\Sigma=\Sigma_3(\mu,M)}.\Biggr.
\ee
Since $\Sigma_3$ is the solution of the stationary equation (12), the second
item in (18) equals to zero. Taking into account (11), we have:
\be
\frac{d\Omega_c(\mu)}{d\mu}=
\frac{\partial V_\mu (\Sigma)}{\partial\mu}
\Biggl|_{\Sigma=\Sigma_3}=
-\frac{N}{3\pi^2}
(\mu^2-\Sigma^2_3)^{3/2}.
\Biggr.
\ee
According to the eq. (15) expression (18) equels to zero, when $\mu\to M_+$.
So the first derivative of the $\Omega (\mu)$ is continious function at the
point $\mu=M$. 
In the following we need the relations, which are consequences of eq. (13):
\[
\frac{d\Sigma_3}{d\mu}\equiv\Sigma'_3=
\Biggl\{
\frac{\partial f_\mu (\Sigma)}{\partial\mu}
\Biggl [
\frac{\partial f_\mu (\Sigma)}{\partial\Sigma}\Biggr]^{-1}
\Biggr\}
\Biggl |_{\Sigma=\Sigma_3}=\Biggr.\nonumber \\
\]
\be
=-\frac{2\sqrt{\mu^2-\Sigma^2_3}}
{\Sigma_3\{\ln
[(\Sigma^2_3+\Lambda^2)/(\mu+\sqrt{\mu^2-\Sigma^2_3}~)^2]
-\Lambda^2/(\Sigma^2_3+\Lambda^2)\}},\nonumber \\[0.2cm]
\ee
\be
\frac{d^2\Sigma_3}{(d\mu)^2}\equiv\Sigma^{''}_3=
-\frac{2(\mu-\Sigma_3\Sigma'_3)}
{\Sigma_4\sqrt{\mu^2-\Sigma^2_3}\{\cdots\}}
+O(\mu^2-\Sigma^2_3).
\ee
Here in the braces in (21) one should mean the same expression as in the braces
in eq. (20). Evidently, that $\Sigma'_3\to 0$ at $\mu\to M$, but the second
derivative $\Sigma^{''}_3$ at $\mu\to M$ turns into $(-\infty)$. Let us now find
second and third derivatives of the TP $\Omega_c(\mu)$. Taking into account
eqs. (20-21) and (19), we obtain
\be
\frac{d^2\Omega_c (\mu)}{(d\mu)^2}=
-\frac{N}{\pi^2}
(\mu^2-\Sigma^2_3)^{1/2}(\mu-\Sigma_3\Sigma'_3)
\ee
\ba
\frac{d^3\Omega_c (\mu)}{(d\mu)^3}&=&
-\frac{N}{\pi^2}
\frac{(\mu-\Sigma_3\Sigma'_3)^2}
{\sqrt{\mu^2-\Sigma^2_3}}
-\frac{N}{\pi^2}\sqrt{\mu^2-\Sigma^2_3}
~(1-\nonumber \\
&-&(\Sigma'_3)^2-\Sigma_3\Sigma_3^{''}).
\ea

Expression (22) at $\mu\to M$ turns into zero, but the third derivative of the TP
$\Omega_c(\mu)$ at $\mu\to M$ turns into infinity.
Hence, at the point $\mu_{2c}=M$ we have discontinuous third derivative of the
thermodynamic potential $\Omega (\mu)$. According to the criterion of the
phase transitions \cite{10}, this mean that at the point $\mu_{2c}(M)$ we have
phase transition of the second order from massive phase B (at $\mu<M$) to
the massive phase С (at $M<\mu<\mu_{1c}(M)$). The main phisical quantity, which
takes different values in those phases, is the density of particles number:
\be
n=-\partial\Omega (\mu)/\partial\mu.
\ee
With the help of eqs. (24) and (17) one can easily show that in the phase В 
the particles density $n_B\equiv 0$. But in the phase С (see (19)) we have:
\be
n_c=\frac{N}{3\pi^2}(\mu^2-\Sigma^2_3
(\mu,M))^{3/2}\not=0.
\ee

Resuming, we should note that in quantum systems with NJL Lagrangians at nonzero
chemical potentials there are three different phases. The first one is massless
chirally symmetric phase A, the second B and the third C are massive with
spontaneous breaking of chiral symmetry phases. The new phase C is not
observed in low dimensional four - fermionic models \cite{9,11}.

In the present paper we have restricted ourselves by condition $M<<\Lambda$.
But in the forthcoming article by Vshivtsev, Zhukovsky and Klimenko the whole
range of variations for the parameter $M$ is considered. 
In this case the same phases A,B and C of the NJL model are possible.
But the critical behaviour of the model is more various. Indeed, we have observed
on the phase diagram $\mu,M$ of the model several tricritical points as well, as
the phase transitions of second and first orders.

\end{document}